\documentclass[aps,prl,twocolumn]{revtex4}
\usepackage{amsmath}
\usepackage{graphicx}
\usepackage[section]{placeins}

\newcommand{\eq}{\begin{equation}}
\newcommand{\fine}{\end{equation}}

\begin{document}

\title{Polarization preserving ultra fast optical  shutter for quantum information
processing}
\author{Nicol\'{o} Spagnolo$^{1}$, Chiara Vitelli$^{1}$, Sandro Giacomini$^{1}$,
Fabio Sciarrino$^{2,1},$ and Francesco De Martini$^{1,3}$ \\
$^{1}$Dipartimento di Fisica dell'Universit\'{a} ''La Sapienza'' and
Consorzio Nazionale Interuniversitario per le Scienze Fisiche della Materia,
Roma 00185, Italy\\
$^{2}$Centro di\ Studi e Ricerche ''Enrico Fermi'', Via Panisperna
89/A,Compendio del Viminale, Roma 00184, Italy\\
$^{3}$ Accademia Nazionale dei Lincei, via della Lungara 10, I-00165 Roma, Italy 
}

\begin{abstract}
We present the realization of a ultra fast shutter for optical
fields, which allows to preserve a generic polarization state,
based on a self-stabilized interferometer. It exhibits high (or
low) transmittivity when turned on (or inactive), while the
fidelity of the polarization state is
 high. The shutter is realized through two
beam displacing prisms and a longitudinal Pockels cell. This can
represent a useful tool for controlling light-atom interfaces in
quantum information processing.
\end{abstract}

\maketitle

In the last few years, quantum information processing (QIP) has attracted a
growing interest. Its optical implementation opens new perspectives both for
quantum communication and quantum computing. Quantum communication is based
on the distribution of photonic entangled states \cite{Gisi02}, while
quantum computing relies on optical gates in the KLM approach\ \cite{Knil02}%
\ or in measurement carried out on complex cluster entangled
states \cite {Raus01}. The previous tasks require the fast
implementation of operation conditioned on single photon
measurements. The active teleportation protocol, which involves
conditional fast-feedforward transformations, has been first
demonstrated by Giacomini et al \cite{Giac02}, and then by
\cite{Ursi02}. Conditional gates have  been also reported by
\cite{Pitt02}. Within the context of one-way quantum computing
feed-forward measurements have been implemented by
\cite{Prev07,Prev07B} and \cite{Vall07}.

In quantum information framework, the ability to perform
fast-switching of an optical field can have different, useful
applications. On one hand, preparation of multi-photon entangled
states by optimized measurements and feed-forward operations can
lead to innovative QIP protocols. On the other one, the coupling
of mesoscopic field and Bose-Einstein condensate has been recently
investigated \cite{Cata07} and requires the implementation of
optical shutters able to switch in a very fast time,while
preserving the quantum state and exhibiting a high extinction
value. Since the content of information is usually encoded in the
polarization degree of freedom of the field, the switching device
should be able to preserve any polarization state of the incoming
radiation. Hence a shutter device based on a fast-pockels cell, as
the one developed by Ref.\cite{Giac02,Prev07,Vall07}, combined
with a polarizing beamsplitter would destroy the carried
information.\\ An alternative solution based on an acousto-optic
modulator requires a longer activation time and leads to an
intensity of the diffracted beam between 0\% and 60\%, while the
zero order contribution is always higher than 15\%. Here we
present the realization of an ultra-fast shutter for optical
field, which preserves a generic polarization state and exhibits a
high transmittivity . The shutter is realized through two beam
displacing prisms and a longitudinal Pockels cell (PC).

\begin{figure}[tbph]
\includegraphics[scale=.35]{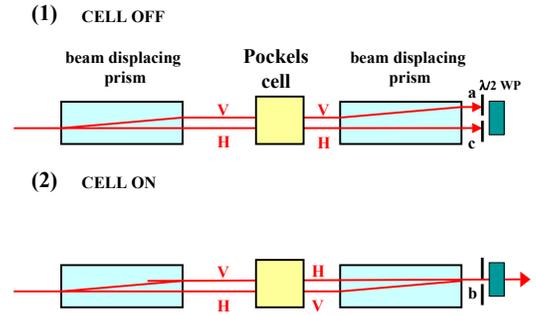} 
\caption{Experimental scheme of the shutter: (1) when the shutter is off the
two beams separated by the two calcites are stopped by the pin-hole (modes $%
\mathbf{a}$\ and $\mathbf{c}$).\ (2) On the contrary when the shutter is on
the two beams are recombined on the second calcite and the resulting beam
(mode $\mathbf{b}$) passes through the pin-hole.}
\label{fig:schema}
\end{figure}
Let us sketch the working details.\ Calcite beam displacing prism
is used to separate an input beam into two orthogonally polarized
output beams. Before passing through a second calcite prism these
are manipulated in a  PC with optical axis oriented at $45^{\circ }$%
(Fig. \ref {fig:schema}). When the PC is off (Fig.
\ref{fig:schema}-(1)) it leaves the polarization state unperturbed
and the beams are further separated in the second calcite .\ In
this situation the shutter is off,\ and  the output beams on modes
$\mathbf{a}$\ and $\mathbf{c}$ are stopped by a pin-hole.\
 On the
other hand,\ when the cell is on,  the PC driving voltage is set
 to induce a $\lambda /2$\ phase shift between the
ordinary and extraordinary components and it can be activated in a
short time ($t<10ns$) by an external TTL signal.\ In this way the
transformation $\overrightarrow{\pi _{H}}\Leftrightarrow
\overrightarrow{\pi _{V}}$\ is implemented. Then the two
orthogonally beams are recombined spatially and temporally in the
second calcite,\ Fig. \ref{fig:schema}-(2). In this situation  the
output field emerges on mode $\mathbf{b}$. At the end of the stage
a $\lambda /2$\ waveplate at $45^{\circ }$\ flips the polarization
of the output beam to restore its initial state. It's worth noting
that the temporal overlap between the two beams is automatically
ensured by the symmetry of the device:\ the $\overrightarrow{\pi
_{H}}$ polarization component of the input beam goes through the
first calcite on a straight path whereas the $\overrightarrow{\pi
_{V}}$\ component's path is deviated.\ At the exit of the first
calcite the two orthogonal polarization components are separated
by a distance $d=4mm$,\ the PC exchanges them and in the second
calcite they are recombined by virtue of the fact that they have
experienced the same overall path deviation. \\ The present device
can be adopted with ultra short pulses ( $200fs$).\ We note that
this system is also stable in phase.\ Indeed the two orthogonally
polarized beams are subjected to the same phase fluctuations since
they propagate along parallel optical paths and share the same
optical mounts.\ The phase difference between the two beams can be
finally controlled by tilting the second calcite \cite{Whit03}.

Let us now analyze the action of the shutter on an input quantum state $%
\left| \varphi \right\rangle $ with generic polarization $\overrightarrow{%
\pi }_{\varphi }=\alpha \overrightarrow{\pi }_{H}+\beta \overrightarrow{\pi }%
_{V},$ where $(\alpha ,\beta )$ are complex numbers satisfying
$\left| \alpha \right| ^{2}+\left| \beta \right| ^{2}=1$ and
$\vec{\pi}_{H}$ and $\vec{\pi}_{V}$ stand for horizontal and
vertical polarization, respectively. The evolution of $\left|
\varphi \right\rangle $ is investigated by looking to the
Heisenberg dynamic of the creation operator associated to the
spatial mode $\mathbf{c}$ with polarization $\overrightarrow{\pi
}_{\varphi }:\widehat{c}_{\varphi
}^{\dagger }=\alpha \widehat{c}_{H}^{\dagger }+\beta \widehat{c}%
_{V}^{\dagger }$. After the first calcite the operator becomes ($e^{i\chi
_{1}}\alpha \widehat{c}_{H}^{\dagger }+e^{i\chi _{2}}\beta \widehat{b}%
_{V}^{\dagger }$), where $\chi_{1}$ and $\chi_{2}$ are the
phase-shifts induced on the two orthogonal polarizations due to
their different optical paths. When the Pockels cell is switched,
the operator evolves into : ($e^{i\chi _{1}}\alpha
\widehat{c}_{V}^{\dagger }+e^{i\chi _{2}}\beta
\widehat{b}_{H}^{\dagger }$), and the output state results after
the
recombination in the second calcite: $e^{i\chi }(\alpha \widehat{b}%
_{V}^{\dagger }+\beta \widehat{b}_{H}^{\dagger })$, where $\chi=\chi_{1}+\chi_{2}$. Finally, after the $%
\frac{\lambda }{2}-$waveplate, we obtain the same polarization state as the
input one $\alpha \widehat{b}_{H}^{\dagger }+\beta \widehat{b}_{V}^{\dagger }
$. On the contrary, if the cell is off, the total operator becomes ($%
e^{i 2 \chi _{1}}\alpha \widehat{c}_{H}^{\dagger }+e^{i 2 \chi _{2}}\beta \widehat{%
a}_{V}^{\dagger }$), and, in this case, the initial polarization state is
lost. We note that this scheme can be adopted in all the visible range by
changing the PC voltage and by exploiting the spectral operating range (from
$350nm$\ to $2.3\mu m$) of the optical grade calcite of the displacers.%
\newline

\begin{figure}[tbph]
\vspace{0.5cm} \includegraphics[scale=0.5]{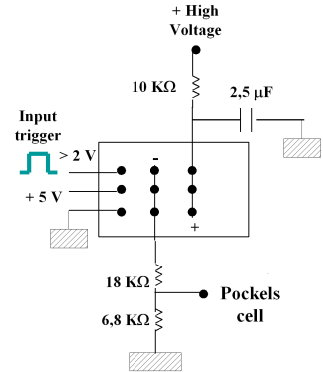}
\caption{Electronic driver of the Pockels cell.\ When the signal
of the trigger is on the PC is activated.}
\label{fig:schema_elettronico}
\end{figure}

\begin{figure}[tbph]
\vspace{0.5cm} \includegraphics[scale=0.4]{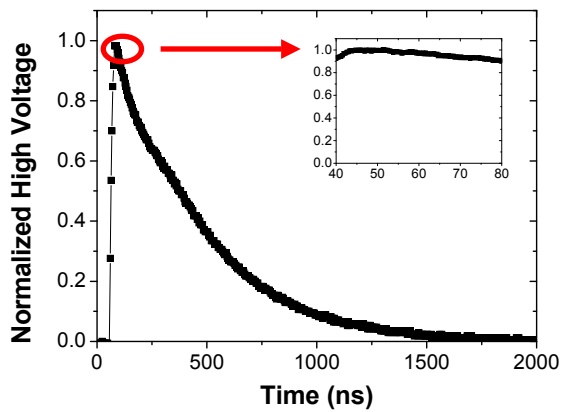}
\caption{Trend of the high voltage signal as function of time.
Inset: the trigger signal remain constant for almost $10ns$. }
\label{fig:trigger}
\end{figure}

The adopted electro optic cell, Lasermetrics  Series 1042, was
composed by the series of two longitudinal PC of same length
$35mm$,\ powered by a high voltage of $3200V$ to produce a
$\lambda /2$\ shift on the incident polarization and driven by the
circuit reported in Fig.\ref{fig:schema_elettronico}. The problem
of realizing a fast electronic circuit transforming a TTL signal
into a calibrated fast pulse in the kV range was solved by a solid
state switch HTS 50-08-UF, characterized by a very low jitter and
a lifetime typical of semiconductor devices. The switch is
triggered by a positive going pulse of 2 to 10 volts amplitude and
generates the signal\ shown in Fig.\ref {fig:trigger}. The pulse
remains constant for a time window of almost $10ns$ and decays
exponentially within $500ns$. The time duration of the driver
pulse has been chosen to satisfy two criteria: (a)\ reduced
low-frequency components and (b) suitable activation time window.
(a) The KD*P crystal suffers the piezoelectric effect: when
excited by a long high voltage pulse an effective coupling is
introduced between the corresponding low frequency spectral
components and the acoustic phononic modes of the crystal. The
corresponding strain causes a mechanical damped oscillation of the
crystal for a time duration longer than the ultra fast activation
time of the shutter.  This effect due to the polarizability of the
Pockels cell is harmonically modulated. Hence, the shutter is
periodically reactivated and several subsequent pulses are
partially transmitted. In order to eliminate this effect an
ultra-short activation pulse is required \cite{Bish06}. (b) The
electronic jitter of the driver circuit, almost $1-2ns$, gives a
lower limit to the time activation window.\newline

\begin{figure}[tbph]
\includegraphics[scale=.35]{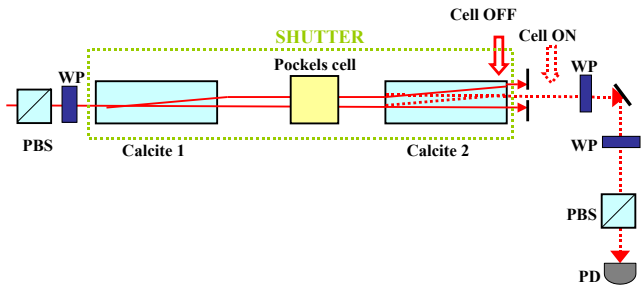}
\caption{Experimental setup.\ A PBS and a $\protect\lambda /2$\ waveplate
allow to vary the polarization of the input beam.\ A second PBS and $\protect%
\lambda /2$ waveplate analyze the polarization state of the output beam
\textbf{b}.\ The signal is detected by a photodiode (PD).}
\label{fig:setup}
\end{figure}

We describe now the experimental characterization of the shutter
device. We used a pulsed laser source centered at $800nm$\ with a
repetition rate of $250kHz$\, and a bandwidth of $1.5nm$, selected
before the shutter by two interferential filters .\ A $\lambda
/2$\ waveplate (WP) and a polarizing beam splitter (PBS) allowed
to vary the polarization of the input beam on the first calcite
(fig. \ref{fig:setup}).\ A second $\frac{\lambda}{2}$-WP and a PBS
realized the polarization analysis of the output beam,\ which was
detected by a photodiode (PD).\

\FloatBarrier

\begin{figure}[tbph]
\includegraphics[scale=0.4]{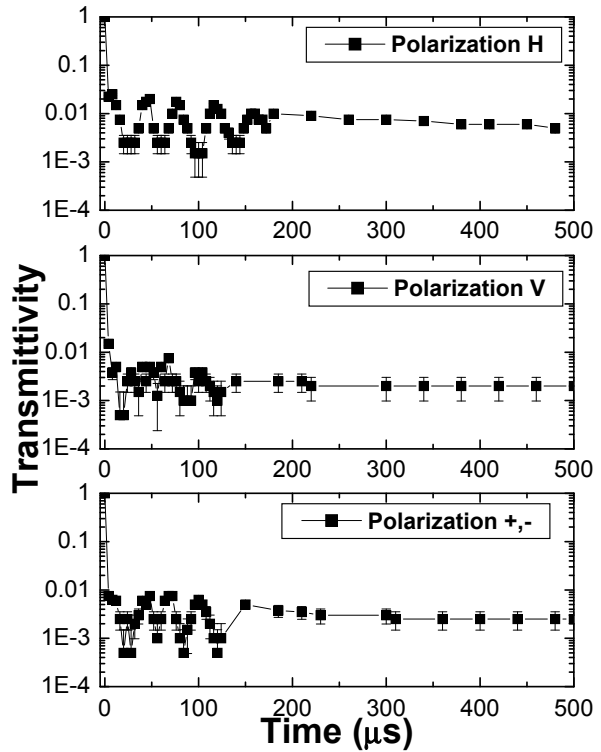}
\caption{Transmittivity for an input state with polarization $\{\protect\pi
_{+},\protect\pi _{-}\}$ and $\{\protect\pi _{H},\protect\pi _{V}\}$
measured with a frequency of the trigger equal to 1 kHz.}
\label{fig:Transmittivity}
\end{figure}

The PC was activated via the circuit above described  (fig. \ref
{fig:schema_elettronico}).\ For different values of the frequency
of the TTL trigger signal, we measured the fidelity
$\mathcal{F}_{ON}$ of the polarization state when the PC was
on,\ the transmittivity $\mathcal{T}_{ON} $ and the
transmittivity $\mathcal{T}_{OFF}$ when the shutter was on and off
respectively:

\begin{eqnarray}
\mathcal{F}_{ON} &=&\frac{I_{i}^{ON}}{I_{i}^{ON}+I_{i^{\perp
}}^{ON}}
\\
\mathcal{T}_{ON} &=&\frac{I_{i}^{ON}+I_{i^{\perp }}^{ON}}{I_{IN}}
\\
\mathcal{T}_{OFF}
&=&\frac{I_{i}^{OFF}+I_{i^{\perp}}^{OFF}}{I_{IN}}
\end{eqnarray}

where $I_{i}^{ON}$ ($I_{i}^{OFF})$\ stands for the measured
intensity on spatial mode $\mathbf{b}$ with polarization state
$\pi_{i}$ equal to the input one when the PC is (is not)
activated.\ $I_{i^{\perp }}^{ON}$\ ($I_{i^{\perp }}^{OFF}$) stands
for the
measured intensity of the analyzed polarization state $\overrightarrow{\pi }%
_{i}^{\perp }$ perpendicular to the input one $\overrightarrow{\pi
}_{i}$.\ $I_{IN}$ stands for the incident intensity on the
shutter. For a trigger signal frequency equal to $1kHz$ we found the following
results:

\begin{flushleft}
\begin{small}
\begin{tabular}{|c|c|c|c|}
\hline
\textbf{Polarization} & $\mathbf{\mathcal{F}_{ON}}$ $\ (\pm 0.001)$ & $%
\mathbf{\mathcal{T}_{ON}}$ $\ (\pm 0.001)$ & $\mathbf{\mathcal{T}_{OFF}}$ $%
\ (\pm 0.001)$ \\ \hline $\overrightarrow{\pi _{+}}$ & 0.956 &
0.991 & 0.0025 \\ \hline $\overrightarrow{\pi _{-}}$ & 0.956 &
0.991 & 0.0025 \\ \hline $\overrightarrow{\pi _{H}}$ & 0.998 &
0.991 & 0.0050 \\ \hline $\overrightarrow{\pi _{V}}$ & 0.998 &
0.991 & 0.0020 \\ \hline
\end{tabular}
\end{small}\end{flushleft}

The transmittivity obtained with the shutter off gives an
estimation of the extinction power of the shutter.\ The mean
transmittivity in this case was $\mathcal{T}_{OFF}=0.003$.\ In
order to verify the absence of the piezoelectric ringing effect we
report in Fig.\ref{fig:Transmittivity} the transmittivity of the
shutter as a function of time. After few $\mu s$ the transmitted
signal is reduced by a factor of $100$, leading to the
transmission of one pulse once activated and the extinction of the
subsequent pulses. When the shutter
was on we obtained a mean fidelity $\mathcal{F}_{ON}=0.998$ for $%
\overrightarrow{\pi _{H}}$\ and $\overrightarrow{\pi _{V}}$\ polarizations,\
and $\mathcal{F}_{ON}=0.956$ for $\overrightarrow{\pi _{+}}$\ and $%
\overrightarrow{\pi _{-}}$ polarizations.\newline

\begin{figure}[t]
\includegraphics[scale=0.8]{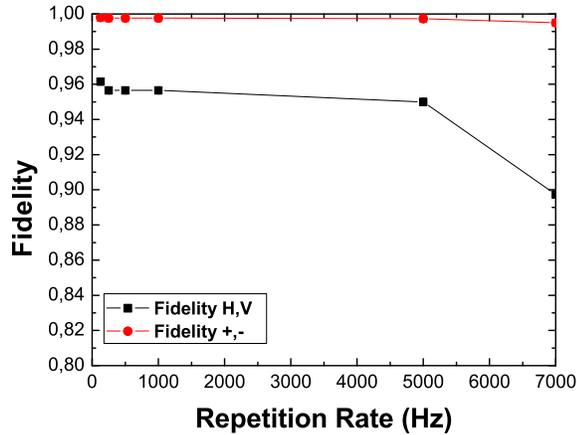}
\caption{Fidelity of the polarization state in the $\{\protect\pi _{+},%
\protect\pi _{-}\}$ and the $\{\protect\pi _{H},\protect\pi _{V}\}$\ basis
versus the frequency of the trigger signal.}
\label{fig:Fidelity}
\end{figure}

We observe at last that the increase of the repetition rate causes
an increase of transmittivity $\mathbf{\mathcal{T}_{OFF}}$ and a
decrease of fidelity $\mathbf{\mathcal{F}_{ON}}$
(fig.\ref{fig:Fidelity}). Indeed for high repetition rate values,
the time interval between two following trigger signals is shorter
than the PC recovery time. By varying the frequency of the trigger
signal, we have at last studied the fidelity
in the two polarization basis: $\{\overrightarrow{\pi _{H}},\overrightarrow{%
\pi _{V}}\}$\ and $\{\overrightarrow{\pi _{+}},\overrightarrow{\pi _{-}}\}$%
.\ We report the experimental results in fig. \ref{fig:Fidelity}. The
fidelity values for the states $(\overrightarrow{\pi }_{+},\overrightarrow{%
\pi }_{-})$ are lower due to the interferometric feature of the
device, however an average fidelity value as high as $97\%$ has
been observed with the present scheme.

In conclusion, we reported the experimental realization and
characterization of a ultra fast  shutter for optical field, based
on a self-stabilized interferometer, which preserves a generic
polarization state with high fidelity and exhibits a high contrast
operation. This device can have direct applications in the context
of measurement induced quantum operations.

We acknowledge support from MIUR\ (PRIN\ 05) and from CNISM\ (Progetto
Innesco 2006).

\end{document}